\documentclass[12pt,aps,prd,superscriptaddress,showpacs,longbibliography,floatfix,nofootinbib]{revtex4-1}

\usepackage[utf8]{inputenc}
\usepackage{slashed}
\pdfoutput=1

\usepackage{color}
\usepackage{graphicx}   % Include figure filed
\usepackage{bm}
\usepackage{amsmath}
\usepackage{amsfonts}
\usepackage{hyperref}
\def\be{\begin{equation}}
\def\ee{\end{equation}}

\begin{document}

\title{A solvable quantum field theory with asymptotic freedom in 3+1 dimensions}

\author{Paul Romatschke}
\affiliation{Department of Physics, University of Colorado, Boulder, Colorado 80309, USA}
\affiliation{Center for Theory of Quantum Matter, University of Colorado, Boulder, Colorado 80309, USA}
%\emailAdd{paul.romatschke@colorado.edu}

\begin{abstract}
  Recently, Ai, Bender and Sarkar gave a prescription on how to obtain $\mathcal{PT}$-symmetric field theory results from an analytic continuation of Hermitian field theories. I perform this analytic continuation for the massless (critical) O(N) model with quartic interaction in 3+1 dimensions. In the large N limit, this theory is exactly solvable, and has negative $\beta$-function in the ultraviolet, and a stable bound state in the infrared. The coupling diverges at a scale $\Lambda_c$, but can be continued into the far infrared. At finite temperature, the theory exhibits two phases separated by a second-order phase transition near $T_c\simeq \Lambda_c/\sqrt{e}$. 
\end{abstract}

\maketitle

\section{Introduction}

Scalar field theory with quartic interaction is one of the best studied quantum field theories in existence. In 3+1 dimensions in the weak coupling limit, its $\beta$-function is positive, suggesting there is a Landau pole in the theory, essentially prohibiting its existence as a continuum interacting quantum field theory. Indeed, quantum triviality for single component scalar field theory with quartic interaction was recently proved \cite{Aizenman:2019yuo}.

So why bother studying this trivial theory?

While the proof for quantum triviality rules out interacting scalar field theory in four dimensions for a single component scalar field, multiple scalar field components are not covered by the proof. This leaves the possibility that replacing a single scalar field by a vector $\phi\rightarrow \vec{\phi}$ with $\vec{\phi}=\left(\phi_1,\phi_2,\ldots,\phi_N\right)$ and $N>1$ exists as an interacting quantum field theory in four dimensions. Indeed, it has long been known that this so-called O(N)-model with N bigger than some critical value $N>N_{\rm crit}$ may exist as an interacting quantum field theory even in \textit{five dimensions} \cite{Parisi:1975im}, which has attracted recent attention, cf. Refs.~\cite{Li:2016wdp,Giombi:2019upv}. In addition, considering multiple components is attractive for a difference reason: if the number of components is taken to be very large $N\gg 1$, the O(N) model possesses an expansion parameter unrelated to the coupling, so that the quantum field theory is exactly solvable in the large N limit \cite{Moshe:2003xn}.

However, for the case of 3+1 space-time dimensions, the large N result for the O(N) model with quartic interaction is unambiguous: the theory has positive $\beta$-function and a Landau pole, cf. \cite{Romatschke:2019gck}. Prevailing knowledge has it that because of the Landau pole, the O(N) model does not exist as a continuum interacting quantum field theory.

So far the story is completely standard, and all the details have been known for many years. However, the story acquires a new twist by incorporating an apparently unrelated development.

In Ref.~\cite{Bender:1998ke}, it was realized that Hermiticity, for many years a given assumption for the existence of quantum mechanics (and hence quantum field theory), may be too strong. Indeed, Ref.~\cite{Bender:1998ke} showed that the spectrum of a Hamiltonian that is non-Hermitian, but invariant under parity $\mathcal{P}$ and time-reversal $\mathcal{T}$, possesses a real, positive and discrete Eigenspectrum.

More recently, Ai, Bender and Sarkar gave a prescription on how to obtain results in $\mathcal{PT}$ symmetric scalar quantum field theory with quartic interaction \cite{Ai:2022csx} via analytic continuation of its ordinary (standard Hermitian) cousin. In a nutshell, the prescription is as follows: given a standard Hermitian field theory with potential $V(\phi)=\lambda \phi^4$ and partition function $Z(\lambda)$, the partition function $Z_{\mathcal{PT}}(g)$ for the $\mathcal{PT}$ symmetric theory, defined by the potential $V(\phi)=-g \phi^4$ can be found via the analytic continuation\footnote{Note that the prescription was conjectured in Ref.~\cite{Ai:2022csx} based on arguments for d=1 and low temperature.}
\be
\label{conj}
\ln Z_{\mathcal{PT}}(g)={\rm Re}\ln Z(\lambda=-g+i 0^+)\,,
\ee
where ${\rm Re}$ denotes the real part. Because of the sign change of the coupling, it was noted already in Ref.~\cite{Bender:1998ke} and references therein that one can expect $\mathcal{PT}$ symmetric analogues of Hermitian theories with positive $\beta$-function to become asymptotically free.

%discuss Bender's result on PT symmetric QED and asymptotic freedom in PT

In this work, I put together large N solvability for the O(N) model and analytic continuation technique to study the non-perturbative properties of $\mathcal{PT}$-symmetric O(N) theory. 

\section{Calculation}

To get started, I define the standard massless Hermitian field theory via the Euclidean path integral
\be
Z(\lambda)=\int {\cal D}\vec{\phi} e^{-\int d^3x \int_0^\beta d\tau \left[\frac{1}{2}\partial_\mu \vec{\phi}\cdot \partial_\mu \vec{\phi}+\frac{\lambda}{N} \left(\vec{\phi}\cdot \vec{\phi}\right)^2\right]}\,,
\ee
where $\vec{\phi}=\left(\phi_1,\phi_2,\ldots,\phi_N\right)$ is an N-component scalar and I have compactified the Euclidean time direction on a thermal circle with radius $\beta=\frac{1}{T}$. It is convenient to perform a Hubbart-Stratonovic transformation of this path integral by introducing the auxiliary field $\zeta$, such that 
\be
Z(\lambda)=\int {\cal D}\vec{\phi} {\cal D}\zeta e^{-\int d^3x \int_0^\beta d\tau \left[\frac{1}{2}\partial_\mu \vec{\phi}\cdot \partial_\mu \vec{\phi}+\frac{i}{2} \zeta \vec{\phi}^2+\frac{N}{16\lambda} \zeta^2\right]}\,.
\ee
One can always split the auxiliary field into a zero-mode and fluctuations, $\zeta=\zeta_0+\zeta^\prime$, and it is well-known that the fluctuations do not contribute to the leading large N partition function.

To leading order in large N, then, the partition function becomes
\be
\lim_{N\gg 1} Z(\lambda)=\int d\zeta_0 \int {\cal D}\vec{\phi}  e^{-\frac{N \beta V \zeta_0^2}{16 \lambda}-\frac{1}{2}\int d^3x d\tau \vec{\phi}\left[-\Box+i \zeta_0\right]\vec{\phi}}\,,
\ee
where $\beta V$ is the space-time volume and $\Box=\partial_\mu\partial_\mu$. Since $\zeta_0$ is independent from ${\bf x},\tau$, the remaining path integral over $\vec{\phi}$ is quadratic and solvable in closed form,
\be
\lim_{N\gg 1} Z(\lambda)=\int d\zeta_0 e^{-\frac{N \beta V \zeta_0^2}{16 \lambda}-\frac{N}{2}{\rm Tr}\ln\left(-\Box+i \zeta_0\right)}\,,
\ee
At finite temperature, the differential operator may be diagonalized using the Matsubara frequencies $\omega_n=2 \pi n T$, $n\in \mathbb{Z}$, so that
\be
\label{Zlambda}
\lim_{N\gg 1} Z(\lambda)=\int d\zeta_0 e^{N \beta V p(\sqrt{i \zeta_0})}\,,
\ee
with the pressure (negative free energy density) per component
\be
p(m)=\frac{m^4}{16\lambda}-\frac{T}{2}\sum_n \int \frac{d^3{\bf k}}{(2\pi)^3}\ln\left(\omega_n^2+{\bf k}^2+m^2\right)\,.
\ee
The thermal sum-integral is standard in TQFT, and in dimensional regularization can be evaluated as \cite{Laine:2016hma,Romatschke:2019gck}
\be
\label{Pm}
p(m)=\frac{m^4}{16\lambda}+\frac{m^4}{64\pi^2}\left(\frac{1}{\varepsilon}+\ln\frac{\bar\mu^2 e^{\frac{3}{2}}}{m^2}\right)+\frac{m^2 T^2}{2\pi^2}\sum_{n=1}^\infty \frac{K_2(n \beta m)}{n^2}\,,
\ee
where $\bar\mu$ is the $\overline{\rm MS}$ renormalization scale and $K_i(x)$ denote modified Bessel functions of the second kind. 

The large N expression (\ref{Zlambda}) for the partition function $Z(\lambda)$ together with (\ref{Pm}) provides the starting point for obtaining the corresponding $\mathcal{PT}$-symmetric partition function $Z_{\mathcal{PT}}(g)$ defined through the Euclidean action
$
S_{\mathcal PT}=\int d^3x \int_0^\beta d\tau \left[\frac{1}{2}\partial_\mu \vec{\phi}\cdot \partial_\mu \vec{\phi}-\frac{g}{N} \left(\vec{\phi}\cdot \vec{\phi}\right)^2\right]$. To this end, I analytically continue the coupling $\lambda\rightarrow -g +i 0^+$ in (\ref{Pm}), obtaining
\be
p(m)=-\frac{m^4}{16 g}+\frac{m^4}{64\pi^2}\left(\frac{1}{\varepsilon}+\ln\frac{\bar\mu^2 e^{\frac{3}{2}}}{m^2}\right)+\frac{m^2 T^2}{2\pi^2}\sum_{n=1}^\infty \frac{K_2(n \beta m)}{n^2}\,.
\ee
The $\mathcal{PT}$-theory coupling $g$ can be non-perturbatively renormalized,
\be
\frac{1}{g_R(\bar\mu)}=\frac{1}{g}-\frac{1}{4\pi^2 \varepsilon}\,,
\ee
so that the renormalized coupling becomes
\be
\label{run}
g_R(\bar\mu)=\frac{4\pi^2}{\ln \frac{\bar\mu^2}{\Lambda_c^2}}\,,
\ee
where $\Lambda_c$ provides the scale of the theory where the coupling diverges.
The $\beta$-function for the $\mathcal{PT}$-theory is 
\be
\beta(g_R)\equiv \frac{\partial g_R(\bar\mu)}{\partial \ln \bar\mu}=-\frac{8 \pi^2}{\ln^2 \frac{\bar\mu^2}{\Lambda_c^2}}=-\frac{g_R^2(\bar\mu)}{2\pi^2}\,,
\ee
which is negative for all real $g_R(\bar\mu)$.
In the large N limit, the running coupling (\ref{run}) and $\beta$-function are exact. The running coupling becomes small in the UV, a hallmark of an asymptotic theory. At $\bar\mu=\Lambda_c$, the coupling diverges, but can be analytically continued into the infrared. Since $g_R(\bar\mu)$ is itself analytically continued from the coupling of the Hermitian field theory, there is no inherent problem with the negative sign of $g_R(\bar\mu<\Lambda_c)$, in fact this regime corresponds to a positive sign for $\lambda$ in the original Hermitian theory.

\begin{figure}
  \centering
  \includegraphics[width=.7\linewidth]{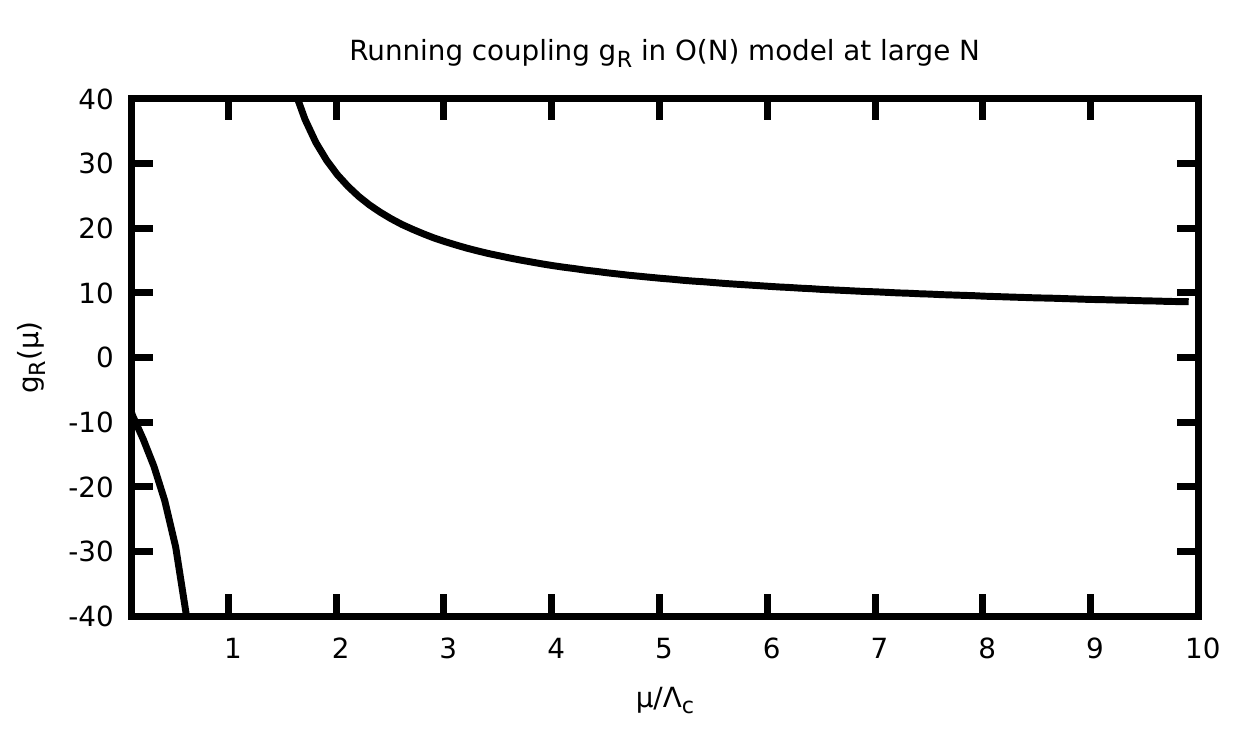}
  \caption{Running coupling for the $\mathcal{PT}$-symmetric theory in the large N limit. \label{fig1}} 
\end{figure}

Using the exact form of the coupling (\ref{run}), the pressure $p(m)$ per component becomes
\be
\label{pfunc}
p(m)=\frac{m^4}{64\pi^2}\ln \frac{(\Lambda_c^2-i 0^+)e^{\frac{3}{2}}}{m^2}+\frac{m^2 T^2}{2\pi^2}\sum_{n=1}^\infty \frac{K_2(n \beta m)}{n^2}\,,
\ee
where I have re-instated the small imaginary part from (\ref{conj}).
In the large N limit, the analytically continued partition function $Z(\lambda=-g+i 0^+)$ can be obtained exactly from (\ref{Zlambda}) using the method of steepest descent
\be
\label{resultZ}
\lim_{N\gg 1} Z(\lambda=-g+i 0^+)=\sum_{i=1}^K e^{N \beta V p(\bar{m}_i)}\,,
\ee
where $\bar{m}_i$, are the locations of all the $K$ saddle points of $p(m)$. The saddle point condition is
\be
\label{saddlecondition}
0=\frac{dp(m)}{dm^2}=\frac{m^2}{32\pi^2}\ln\frac{(\Lambda_c^2- i0^+) e^1}{m^2}-\frac{m T}{4\pi^2} \sum_{n=1}^\infty \frac{K_1(n \beta m)}{n}\,.
\ee

At zero temperature (deep in the infrared), there are two solutions to (\ref{saddlecondition}) given by
\be
\label{gapsol1}
m_0(T=0)=0\,,\quad m_1(T=0)=\sqrt{e}\Lambda_c\,.
\ee
The first solution corresponds to the start of usual perturbation theory, whereas in perturbative approaches, the second one is usually dismissed. Treating both solutions on equal footing, one finds
\be
\label{T0}
\lim_{N\gg 1} Z(\lambda=-g,T=0)=e^{\frac{N \beta V \Lambda_c^4 e^2}{128 \pi^2}}+1\,,
\ee
where the contribution from $m=m_0$ is subleading at large N. (This seems very similar to a situation of two competing thermodynamic phases with different partial pressures -- the phase with the larger partial pressure is thermodynamically favored.) From (\ref{conj}), (\ref{T0}) one therefore obtains the partition function $Z_{\mathcal{PT}}(g)$ for the $\mathcal{PT}$ symmetric theory. As a consequence, one finds for the pressure per component of the $\mathcal{PT}$ symmetric theory in dimensional regularization
\be
p_{\mathcal{PT}}(T=0)= \frac{\Lambda_c^4 e^2}{128 \pi^2}\,.
\ee

At small but finite temperature T, a numerical solution to (\ref{saddlecondition}) still reveals two real-valued solutions $m=\bar{m}$. The analytically continued partition function $\lim_{N\gg 1} Z(\lambda=-g+i 0^+,T=0)$ is the sum of the exponents of the pressure function $p(m)$ evaluated at these saddles, which are also real-valued. In the large N limit, the maximal value of $p(\bar{m}_i)$ dominates the partition function so that $p_{\mathcal{PT}}(T)={\rm max}\left(p(\bar{m}_i)\right)$. 

\begin{figure}
  \centering
  \includegraphics[width=.7\linewidth]{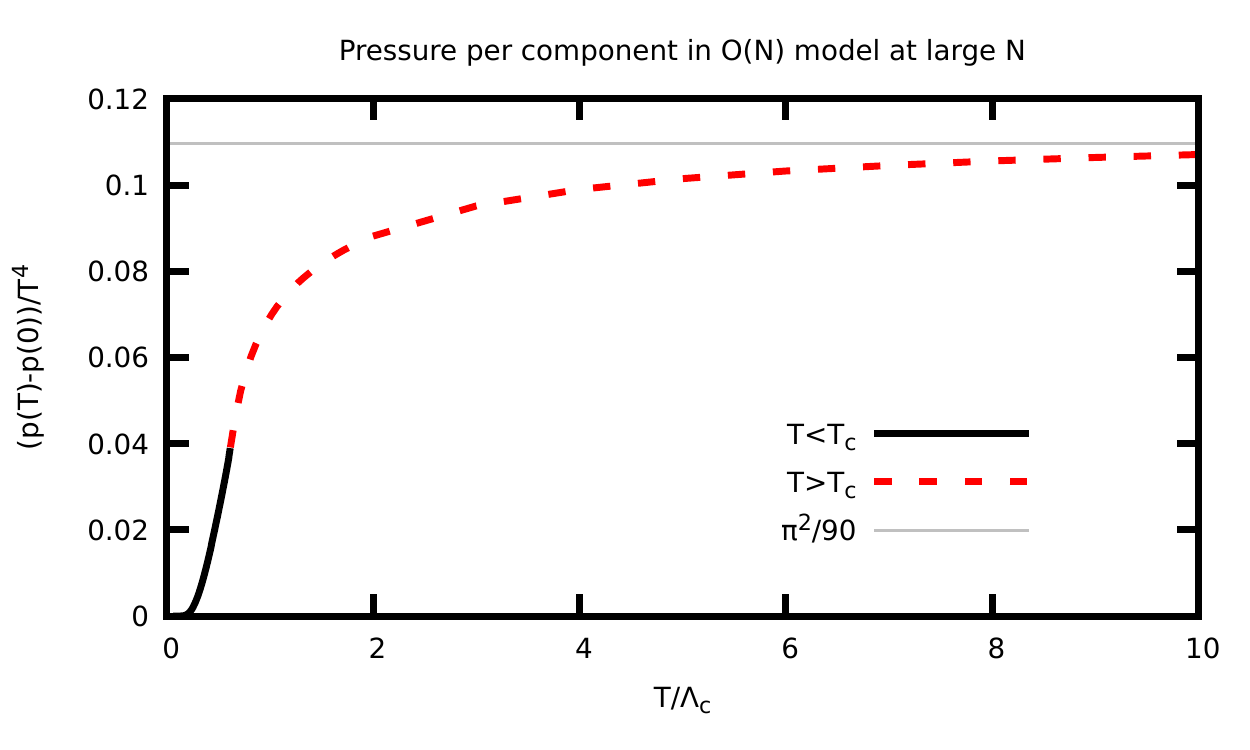}
  \caption{Pressure per component in the $\mathcal{PT}$ symmetric theory as a function of temperature. Different line styles where used for the low temperature and high temperature phase. For reference, the value for a single bosonic degree of freedom, corresponding to $p=\frac{\pi^2 T^4}{90}$ is indicated by a thin line. \label{fig2}} 
\end{figure}

Above a certain temperature $T_c\simeq 0.616 \Lambda_c$, solutions to (\ref{saddlecondition}) are no longer real-valued in $m$. Instead, there is a pair of complex solutions $m_\pm$ with $m_-=m_+^*$ the complex conjugate of $m_+$ with associated complex-valued pressures $p(m_-)=p^*(m_+)$. Only the presence of the small regulator $0^+$ in (\ref{saddlecondition}) breaks this symmetry, such that for non-vanishing $0^+$ the real part of the pressure is larger for one saddle ($m_+$) than for the other saddle. For any non-vanishing $0^+$ then, in the large N limit the saddle with the larger real part of the pressure dominates over the other so that one finds for the pressure per component of the $\mathcal{PT}$ symmetric theory in dimensional regularization
\be
\label{FR}
p_{\mathcal{PT}}(T>T_c)= {\rm Re}\, p(m_+)\,,
\ee
with $p(m)$ given by (\ref{pfunc}). After this result, the limit $0^+\rightarrow 0$ may be taken without further problems.

I acknowledge that it is not exactly elegant that obtaining (\ref{FR}) requires keeping a finite regulator $0^+$ until the very end. This part of the calculation would clearly benefit from a more rigorous treatment using for instance Lefshetz thimbles \cite{Witten:2010cx}.

A plot of $p_{\mathcal{PT}}(T)$ is shown in Fig.~\ref{fig2}. From this figure, it can be seen that $p_{\mathcal{PT}}(T)$ appears well-behaved thermodynamically, and apparently approaching the free theory limit of one bosonic degree of freedom for high temperature. (Note that $p_{\mathcal{PT}}(T)$ may overshoot the free theory limit for sufficiently high temperature. This does not seem to be inconsistent with the perturbative expectation when the coupling constant is negative).

\begin{figure}
  \centering
  \includegraphics[width=.7\linewidth]{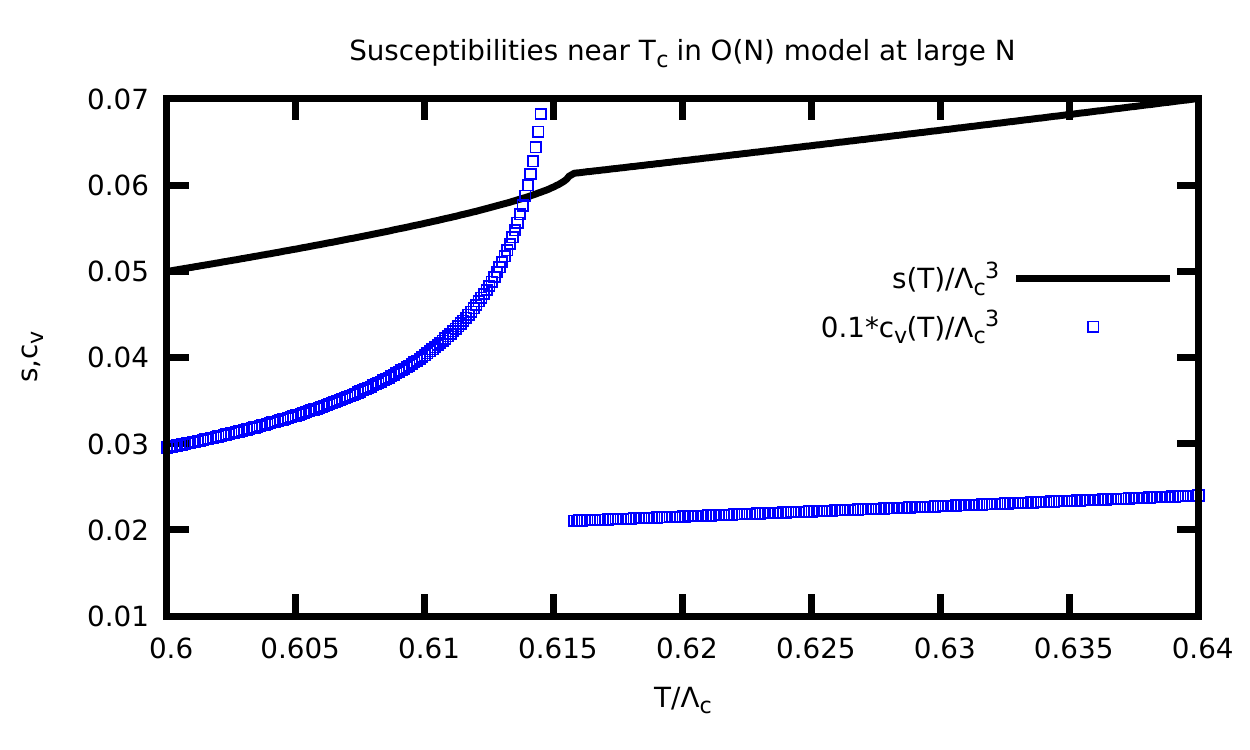}
  \caption{Entropy density $s(T)$ and specific heat $c_v(T)$ per component in the $\mathcal{PT}$ symmetric theory as a function of temperature near $T=T_c$. One can clearly see the jump in the specific heat, indicating a second-order phase transition at $T=T_c\simeq 0.616 \Lambda_c$. \label{fig3}} 
\end{figure}

Near $T=T_c$, I evaluate the entropy density $s=\frac{dp}{dT}$, shown in Fig.~\ref{fig3}. The entropy density is found to be continuous, but exhibiting a cusp at $T=T_c$. The specific heat $c_v(T)=T \frac{d^2p}{dT^2}$ is discontinuous as a function of temperature, thus signaling a second-order phase transition at $T=T_c$. This situation seems to be very similar to the large N Gross-Neveu model in 2+1 dimensions, which in the massless case shows a second-order phase transition that is softened to a cross-over when finite quark masses are incorporated. Perhaps a similar phenomenon will occur for the $\mathcal{PT}$ symmetric O(N) model away from criticality.

Finally, one may ask whether bound states occur in the low temperature phase. To this end, it is necessary to re-instate fluctuations $\zeta^\prime$ in the path integral. This way, one can calculate the Euclidean Green's function $D(x)$ for the auxiliary field $\zeta$ at large N, finding (cf. \cite{Weiner:2022kgx})
\be
D_{\mathcal{PT}}(k)=\frac{1}{-\frac{N}{8(g-i 0^+)}+N \Pi(k)}\,,
\ee
with $\Pi(k)$ at zero temperature given by
\be
\Pi(k)=\frac{1}{2}\int\frac{d^4p}{(2\pi)^4}\frac{1}{(p-k)^2+m^2}\frac{1}{p^2+m^2}\,,
\ee
and $m=m_1$ the solution to the zero temperature gap equation (\ref{gapsol1}).
Evaluating $\Pi(k)$, one finds that the same non-perturbative renormalization also renders $D_{\mathcal{PT}}^{-1}(k)$ finite, so that using the running coupling (\ref{run}), one finds
\be
D^{-1}_{\mathcal{PT}}(k)=\frac{N}{32\pi^2}\left[ \ln \frac{\Lambda_c^2 e^2}{m^2}-
  2\sqrt{\frac{k^2+4m^2}{k^2}}{\rm atanh}\sqrt{\frac{k^2}{k^2+4m^2}}\right]\,.
\ee
Upon analytically continuing the Euclidean momentum to real frequencies $k_4\rightarrow i \omega-0^+$ and numerically solving the dispersion relation $D^{-1}_{\mathcal{PT}}(k)=0$, one finds a stable bound state with a mass of 
\be
m_\zeta \simeq 1.84 m_1\simeq 3 \Lambda_c\,.
\ee

\section{Summary and Conclusion}

In this work, I have studied the $\mathcal{PT}$ symmetric massless O(N) model in the large N limit using an analytic continuation to negative coupling from the original Hermitian quantum field theory. I find that the $\mathcal{PT}$ symmetric massless O(N) model is non-perturbatively renormalizable at large N, exhibiting asymptotic freedom. The scale of the theory is set by the point where the coupling diverges in the infrared, at $\bar\mu=\Lambda_c$. However, the theory may be analytically continued into the far infrared.

At finite temperature, the $\mathcal{PT}$ symmetric massless O(N) model seems to posses two phases, a high-temperature phase with pressure proportional to the number of degrees of freedom $N$, and a low-temperature phase, separated by a second-order phase transition near $T_c\simeq 0.616 \Lambda_c$. I showed that there exists a stable bound state of pairs $\vec{\phi}\cdot\vec{\phi}$ in the low temperature phase with a mass of $m_\zeta\simeq 3 \Lambda_c$.

Let's recall now the original Hermitian O(N) model with partition function (\ref{Zlambda}). It's running coupling is given by
\be
\lambda_{R}(\bar\mu)=\frac{4\pi^2}{\ln \frac{\Lambda_{LP}^2}{\bar\mu^2}}\,,
\ee
where $\bar\mu=\Lambda_{LP}$ is the location of the Landau pole of the theory. But because of the analytic continuation $\lambda\rightarrow - g$, this implies that the $\mathcal{PT}$-symmetric theory \textit{is the same} as the standard Hermitian theory. In particular, the Landau pole in the Hermititan O(N) model is the same as the scale $\Lambda_c$ in the $\mathcal{PT}$-symmetric theory. From the point of view of the $\mathcal{PT}$-symmetric theory, the scale $\Lambda_c$ seems to be benign, only indicating the presence of two distinct phases of the theory.

In the prevailing interpretation, the Hermitian O(N) model cannot be continued above $\bar\mu>\Lambda_{LP}$ because the renormalized coupling changes sign. Using the $\mathcal{PT}$-symmetric theory to interpret the negative coupling regime, this prevailing interpretation may be questioned. In particular, the insights from $\mathcal{PT}$-symmetric field theory allow for the analytic continuation for momentum scales beyond Landau pole. For the  O(N) model at large N, this analytic continuation seem to give results that are physically sensible.

Many aspects of the $\mathcal{PT}$ symmetric O(N) model can and should be studied, and it seems to me that the model appears sufficiently interesting to warrant such investigations in the future. 

\section*{Acknowledgments}

I would like to thank Carl Bender for his stimulating theory colloquium on $\mathcal{PT}$-symmetric field theory at Arizona State University, and writing inspiring articles on interesting subjects over many years. Furthermore, I would like to thank the outstanding theory students and postdocs at CU Boulder, in particular Seth Grable, Scott Lawrence, Ryan Weller and Max Weiner, for many illuminating discussions. Finally, I would like to thank E.~Pr\'eau and W.~Ai for helpful comments on the manuscript. This work was supported by the Department of Energy, DOE award No DE-SC0017905. 

\bibliography{PT}
\end{document}